\documentclass[conference]{IEEEtran}

\IEEEoverridecommandlockouts

\usepackage[pdftex]{graphicx}
\usepackage[cmex10]{amsmath,empheq}
\usepackage{cite}
\usepackage{url}
\usepackage{color}
\usepackage{microtype}
\usepackage{amssymb}
\usepackage{relsize}
\usepackage{lipsum}
\usepackage{epstopdf}
\usepackage{enumitem}
\usepackage{bm}
\usepackage{nicefrac}
\usepackage[percent]{overpic}
\usepackage{subfig}
\usepackage{multirow}

\begin{document}

\title{DeepRx MIMO: Convolutional MIMO Detection with Learned Multiplicative Transformations\vspace{-3mm}}

\newcommand{\todo}[1]{}
\renewcommand{\todo}[1]{{\color{red}{#1}}\PackageWarning{TODO:}{#1!}}
\newcommand{\changed}[1]{{\color{blue}{#1}}}

\author{
\IEEEauthorblockN{Dani Korpi, Mikko Honkala, and Janne M.J. Huttunen}
\IEEEauthorblockA{\textit{Nokia Bell Labs}\\
\textit{Espoo, Finland \vspace{-10mm}}}
\and
\IEEEauthorblockN{Vesa Starck}
\IEEEauthorblockA{\textit{Nokia Mobile Networks}\\
\textit{Espoo, Finland \vspace{-10mm}}}
}

\maketitle

\begin{abstract}
Recently, deep learning has been proposed as a potential technique for improving the physical layer performance of radio receivers. Despite the large amount of encouraging results, most works have not considered spatial multiplexing in the context of multiple-input and multiple-output (MIMO) receivers. In this paper, we present a deep learning-based MIMO receiver architecture that consists of a ResNet-based convolutional neural network, also known as DeepRx, combined with a so-called transformation layer, all trained together. We propose two novel alternatives for the transformation layer: a maximal ratio combining-based transformation, or a fully learned transformation. The former relies more on expert knowledge, while the latter utilizes learned multiplicative layers. Both proposed transformation layers are shown to clearly outperform the conventional baseline receiver, especially with sparse pilot configurations. To the best of our knowledge, these are some of the first results showing such high performance for a fully learned MIMO receiver.
\vspace{-1mm}
\end{abstract}

\section{Introduction}
\vspace{-1mm}

Implementing digital radio functionality with neural networks is one of the emerging concepts in the field of wireless communications. Such neural networks allow fast and efficient implementation of the receiver using neural network chips/AI accelerators. It is also likely that under some circumstances learning-based solutions will result in higher performance, for example, under particular channel conditions, high user equipment (UE) mobility, and/or with very sparse reference signal configurations.

In our earlier work \cite{honkala2020}, we have considered machine learning (ML)-based physical layer receiver (RX) processing. We developed a convolutional neural network (CNN) architecture referred to as DeepRx, which carries out channel estimation, equalization, and demapping jointly. We showed that DeepRx can outperform conventional radio receivers in single-input and multiple-output (SIMO) scenarios where several RX antennas are receiving individual transmissions without spatial multiplexing.

In this paper, we extend our DeepRx architecture to facilitate multiple-input and multiple-output (MIMO) detection, which requires separation of the multiple overlapping spatial streams during the equalization and symbol detection phase. To this end, we propose two novel trainable neural architectures, which are prepended to the beginning of the original DeepRx architecture. These processing techniques introduce expert knowledge to the ML-based DeepRx and consequently simplify the learning task. This is shown to result in significantly higher performance, the ensuing MIMO-compatible DeepRx clearly outperforming convention linear minimum mean square error (LMMSE)-based benchmark receivers.

\subsection{Related work}

As mentioned, recently there has been growing interest into applying different ML techniques in radio physical layer processing. For instance, fully learned receivers for single-input single-output (SISO)/SIMO scenarios have been studied, e.g., in \cite{ye18,zhao2018} in addition to our own earlier work in \cite{honkala2020}. There are also more extreme approaches which aim at learning the complete end-to-end link from the transmitter to the receiver \cite{oshea17,aoudia20}. For a more comprehensive list of prior art, see \cite{honkala2020} and the references therein.

For MIMO, a predominant method has been to utilize ML in some type of model-based approaches. In \cite{Samuel19a}, a so-called detection network (DetNet) is proposed, which learns to perform iterative detection based on certain compressed sufficient statistics. In particular, the method utilizes the empirical channel correlation matrix and a matrix multiplication between the transposed channel matrix and received signal vector. This is linked to our work as we will show that only the latter is necessary for efficient MIMO detection, although our results indicate that the best approach is to allow the neural network to learn everything from data. In addition to this, \cite{pratik20} proposed a neural network based iterative (recurrent) decoding algorithm, operating in conjunction with a classical channel estimation and interpolation algorithm.

\section{System Model}

In this paper, we consider a MIMO orthogonal frequency-division multiplexing (OFDM) system with $N_T$ layers or spatial streams and $N_R$ RX antennas. For simplicity, it is assumed that the number of spatial streams is equal to the number of transmit (TX) antennas. Let us denote the number of OFDM symbols within a transmission time interval (TTI) by $S$ (e.g., typically $S = 14$ in 5G) and the number of utilized subcarriers by $F$. With this, the total received signal after the fast Fourier transform (FFT) can be expressed as
\begin{align}
\mathbf{y}_{ij} = \mathbf{H}_{ij} \textbf{x}_{ij} + \mathbf{n}_{ij}
  \label{eq:rx_signal}, 
\end{align}
where $i$ and $j$ denote the subcarrier and OFDM symbol indices, respectively, $\mathbf{y}_{ij} \in \mathbb{C}^{N_R \times 1}$ and $\textbf{x}_{ij} \in \mathbb{C}^{N_T \times 1}$ are the received and transmitted signals, respectively, $\mathbf{H}_{ij} \in \mathbb{C}^{N_R \times N_T}$ is the channel  over the $i$th subcarrier in the $j$th OFDM symbol, and $\mathbf{n}_{ij} \in \mathbb{C}^{N_R \times 1}$ is the noise-plus-interference signal. The last term incorporates also any inter-carrier- or inter-symbol-interference, which might be caused by the actual physical time-domain channel (note that $\mathbf{H}_{ij}$ can only describe how the temporal channel appears after the FFT).

\subsection{Calculating the Raw Channel Estimate}

The first phase in the RX processing is to calculate the raw channel estimate using the demodulation reference signals (DMRS), also referred to as pilots. In MIMO transmissions, each layer has its own pilots, which are separated from the pilots of other layers in frequency, time, and/or code domain. Regardless of the means of pilot multiplexing, the first raw channel estimate is calculated as
\begin{align}
\widehat{\mathbf{H}}_{ij,\text{raw}} = \mathbf{y}_{ij} \mathbf{x}_{ij}^H, \quad (i,j) \in \mathcal{P}   \label{eq:raw_channel_estimate}
\end{align}
where $\mathcal{P}$ denotes the set of indices corresponding to pilot locations in the TTI time-frequency grid, and $(\cdot)^H$ denotes the Hermitian transpose.

If there is no code-domain multiplexing, i.e., the pilots are orthogonal, the final raw estimate is simply $\widehat{\mathbf{H}}_{ij} = \widehat{\mathbf{H}}_{ij,\text{raw}}$. However, since there is only a limited amount of orthogonal pilot patterns, 5G supports the use of code-division multiplexing (CDM) to provide separable pilots for all layers. In this work, a CDM group size of 2 in the frequency domain is assumed (FD-CDM2), meaning that the pilots of two layers are overlapping in time and in frequency. Then, the final raw estimate $\widehat{\mathbf{H}}_{ij}$ is obtained by averaging the initial raw estimates of subsequent pilot symbols in the frequency domain within each CDM group. Note that the resulting CDM channel estimate has frequency indices that are in between the original pilot indices in $\mathcal{P}$, which must be considered in the interpolation phase.

\subsection{LMMSE-Based Receiver}
\label{sec:lmmse}

In this work, the widely-used LMMSE-based receiver is used as the benchmark. First, the raw channel estimate $\widehat{\mathbf{H}}_{ij}$ is interpolated and filtered to provide a channel estimate for all $(i,j)\in \mathcal{D}$, where $\mathcal{D}$ denotes the set of data-carrying resource element (RE) indices. In this work, the benchmark receiver utilizes splines in the frequency-domain interpolation, while linear interpolation is used in the time direction. Having obtained the full channel estimate, the equalized symbols are then calculated as:
\begin{align}
\widehat{\mathbf{x}}_{ij} = \left(\widehat{\mathbf{H}}_{ij}^H \widehat{\mathbf{H}}_{ij} + \hat{\sigma}^2_n \mathbf{I}\right)^{-1} \widehat{\mathbf{H}}_{ij}^H \mathbf{y}_{ij},\quad (i,j)\in \mathcal{D}, \label{eq:mmse_eq}
\end{align}
where $\hat{\sigma}^2_n$ is the noise power estimate and $\mathbf{I}$ is a $N_T \times N_T$ identity matrix.

The soft bits are then obtained by calculating the log-likelihood ratios (LLRs) based on $\widehat{\mathbf{x}}_{ij}$. In this work, the so-called max-log-MAP demapper is used, which is a widely used approximation of the optimal log maximum a-posteriori demapping rule. For further details about the demapper, please refer to \cite{honkala2020,shental2019}. The final stage of the receiver is to feed the LLRs through the low-density parity-check (LDPC) decoder, which processes the LLRs to provide the final information bits.

\section{DeepRx MIMO Receiver with Expert Knowledge-Based Transformation Layer}

In our previous work \cite{honkala2020}, we introduced a CNN ResNet architecture for a SIMO receiver, referred to as DeepRx. The input to DeepRx is essentially formed by combining the (post-FFT) received data ($\mathbf{y}_{ij}$) and the raw channel estimate ($\widehat{\mathbf{H}}_{ij}$) over a single TTI, and feeding them through multiple layers of ResNet blocks with depthwise separable convolutions and ReLu activations (in some variations also the reference pilot symbols ($\mathbf{x}_{ij}, (i,j) \in \mathcal{P}$) can be fed to DeepRx). The output of DeepRx consists of LLRs for all of the bits within all REs over the TTI. Further details are found in \cite{honkala2020}.

When extending the aforementioned type of DeepRx architecture to MIMO, the input and output arrays are defined~as:
\begin{itemize}
\item Received data signals are denoted by $\mathbf{Y} \in \mathbb{C}^{F \times S \times  N_R}$.
\item The raw-channel estimate is denoted by $\widehat{\mathbf{H}} \in \mathbb{C}^{F \times S \times N_R \times N_T}$, where each $N_R \times N_T$ subarray is the final raw estimate $\widehat{\mathbf{H}}_{ij}$ for the corresponding RE. As opposed to the benchmark receiver, DeepRx uses the very simple nearest neighbor interpolation, meaning that the channel estimate of each data RE $(i,j) \in \mathcal{D}$ is selected based on the raw estimate of the nearest pilot-carrying RE $(i,j) \in \mathcal{P}$.
\item The output of the MIMO DeepRx is the array $\mathbf{L} \in \mathbb{C}^{F \times S \times N_T \times N_\textrm{b}}$  where $N_\textrm{b}$ is the number of bits (the maximum number of bits if modulation is varied).
\end{itemize}

The naive MIMO solution is to reshape the channel estimate to a $F \times S \times N_R N_T$ array, concatenate it with the received data signal array, and feed the resulting array directly to a DeepRx type network \cite{honkala2020}. However, based on our experiments,  such architectures are only able to reach mediocre performance with high-order MIMO scenarios, falling far behind the exceptionally high performance DeepRx achieves in SIMO scenarios. Considering also that separating different MIMO layers is an order of magnitude more demanding problem than signal detection without spatial multiplexing, it could perhaps be possible to improve performance with significantly expanded architectures (e.g., a larger number of ResNet blocks and/or channels\footnote{In the context of neural networks, the term channel refers to the number of CNN channels.}). However, training such a large network would require enormous amount of computational resources and would also lead to very inefficient inference.

\begin{figure*}
  \centering
  \includegraphics[trim=0px 250px 40px 0px, clip,width=0.9\textwidth]{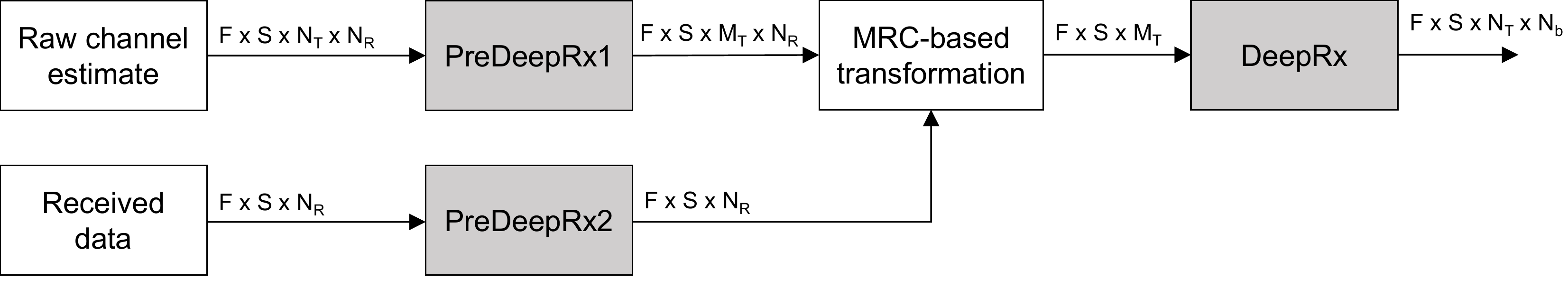}
  \caption{The MRC-based transformation for MIMO DeepRx, where the gray blocks include trainable weights.}
	\label{fig:MRC}
	\vspace{-4mm}
\end{figure*}

To help DeepRx learn the complex operations involved in MIMO detection with significantly reduced computational complexity, we propose two novel transformation layers that can be combined with any neural network receiver. Both of them are incorporated to the input of the primary neural network receiver, such as DeepRx, and trained together with the primary receiver. The two alternative transformations are
\begin{enumerate}
\item Maximum ratio combining (MRC)-based preprocessing using learned virtual spatial streams (Fig.~\ref{fig:MRC}),
\item Fully learned multiplicative transformation/preprocessing (Fig.~\ref{fig:mult_layer}).
\end{enumerate}
The reasoning behind proposing such transformation stems from the fact that a neural network cannot natively perform multiplications between its inputs. Considering that many receiver operations, when carried out in the traditional manner, involve such multiplications, it is reasonable to assume that providing DeepRx with ways to perform multiplicative operations between inputs can improve convergence and final performance. In addition, these transformations scale rather favorably with respect to the number of RX antennas, reducing the pressure to increase the size of the primary neural network receiver for larger RX arrays. Below, we describe these alternatives in detail.

\subsection{MRC-Based Transformation}

The first option for preprocessing the data is based on the so-called MRC, which can be considered as a type of partial equalization invoking an (invalid) assumption that all the spatial streams experience fully orthogonal channel realizations. In the simplest case, where MRC is applied directly on the received signal, the MRC transformation matrix is given by 
\begin{align}
\mathbf{G}_{ij}=\mathbf{S}_{ij} \widehat{\mathbf{H}}_{ij}^H \in \mathbb{C}^{N_T \times N_R}, \label{eq:basic_mrc}
\end{align}
where $\widehat{\mathbf{H}}_{ij}\in\mathbb{C}^{N_R\times N_T}$ is again the subarray of $\widehat{\mathbf{H}}$ corresponding to the $ij$th RE, and
\begin{align}
  \mathbf{S}_{ij}=\textrm{diag}( \Vert
\widehat{\mathbf{h}}_{ij,1}\Vert^{-2},\ldots, \Vert
\widehat{\mathbf{h}}_{ij,N_T}\Vert^{-2}),
\end{align}
with $\widehat{\mathbf{h}}_{ij,k}$ denoting the $k$th column of $\widehat{\mathbf{H}}_{ij}$.  The transformed output is obtained via a vector-matrix multiplication as $\mathbf{y}_{ij,\textrm{MRC}}=\mathbf{G}_{ij} \mathbf{y}_{ij}$, where $\mathbf{y}_{ij} \in \mathbb{C}^{N_R \times 1}$ is the $ij$th RX signal vector (extracted from $\mathbf{Y}$). The output of the MRC transformation, operating on the physical input signal and channel estimate, corresponds to the transmitted spatial streams, although the equalization is still very incomplete. For this reason, the MRC transformation only serves as a preprocessing stage, which must still be fed to the DeepRx to detect the bits.

In this work, we generalize the MRC transformation such that its output dimension can be freely adjusted. The overall receiver architecture utilizing such an MRC-based transformation is presented in Fig.~\ref{fig:MRC}. Instead of directly feeding the received data to the MRC block, the MRC is preceded by a so-called PreDeepRx network, consisting of separate neural networks for the received signal and the channel estimate. Therefore, the input to the MRC-processing consists of the following components:
\begin{align}
  \widetilde{\mathbf{H}}&= f_\textrm{PreDeepRx1}\left(\widehat{\mathbf{H}}\right) \in \mathbb{C}^{ F \times S\times M_T \times N_R}\nonumber\\
  \widetilde{\mathbf{Y}} &= f_\textrm{PreDeepRx2}\left(\mathbf{Y}\right) \in \mathbb{C}^{F\times S \times  N_R}
\end{align}
where $f_\textrm{PreDeepRx1}$ and $f_\textrm{PreDeepRx2}$ are complex-valued 3-block ResNets with 3x3 filters, and $M_T \geq N_T$ is the number of virtual spatial streams. This means that the PreDeepRx1 network augments the channel to represent $M_T$ virtual spatial streams observed over the $N_R$ antennas, while PreDeepRx2 processes only the RX signal.

Using the extended channel $\widetilde{\mathbf{H}}$, the MRC transformation matrix $\widetilde{\mathbf{G}}_{ij} \in \mathbb{C}^{M_T \times N_R}$ is formed similar to \eqref{eq:basic_mrc}, and the generalized MRC transformation output is given by
\begin{align}
\tilde{\mathbf{y}}_{ij,\mathrm{MRC}} = \widetilde{\mathbf{G}}_{ij} \tilde{\mathbf{y}}_{ij} \in \mathbb{C}^{M_T \times 1},
\end{align}
where $\tilde{\mathbf{y}}_{ij} \in \mathbb{C}^{N_R \times 1}$ is the $ij$th subarray of $\widetilde{\mathbf{Y}}$. By repeating the transformation for all REs, the full transformed array is obtained as $\widetilde{\mathbf{Y}}_\mathrm{MRC} \in \mathbb{C}^{F \times S \times M_T}$, which represents the input of the primary DeepRx network.
  
\subsection{Fully Learned Multiplicative Transformation}

Another method for improving the performance of the MIMO DeepRx is to introduce learnable multiplication blocks. The motivation for this is the observation that the MRC-based preprocessing brings expert knowledge into the overall system via the matrix multiplication between the raw channel estimate and received MIMO signal. A logical evolution of this is to limit the expert knowledge only to the concept of multiplication between inputs, but abstain from imposing any other assumptions regarding the processing flow.

\begin{figure*}
  \centering
  \includegraphics[trim=0px 165px 11px 0px, clip,width=0.9\textwidth]{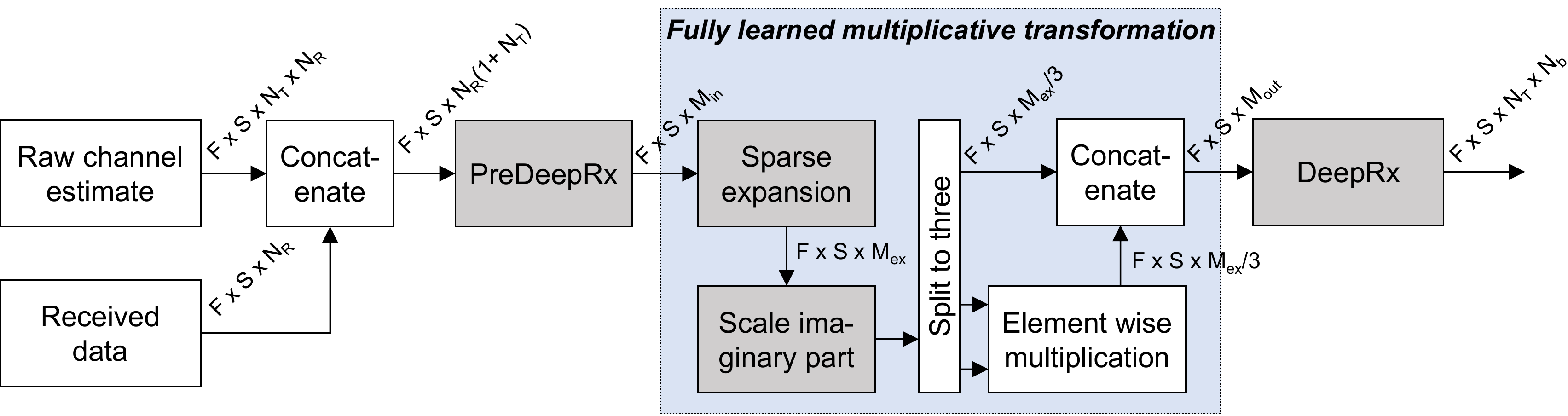}
  \caption{MIMO DeepRx with fully learned multiplicative transformation. Gray blocks include trainable weights.}
	\label{fig:mult_layer}
	\vspace{-4mm}	
\end{figure*}

Figure~\ref{fig:mult_layer} presents a MIMO DeepRx architecture utilizing such fully learned multiplicative preprocessing. The learned multiplicative layer allows the neural network to learn which inputs or channels to multiply before feeding them to the primary DeepRx part. Similar to the MRC-based preprocessing, the input data arrays are first fed through a ResNet referred to as PreDeepRx. This phase can be expressed as
\begin{align}
  \widetilde{\mathbf{Z}} = f_\textrm{PreDeepRx}\left(\mathbf{Y}, \widehat{\mathbf{H}}\right) \in \mathbb{C}^{F\times S \times  M_\mathrm{in}},
\end{align}
where $f_\textrm{PreDeepRx}$ is a complex-valued 3-block ResNet with 3x3 filters and $M_\mathrm{in}$ is the number of channels at its output, which are consequently fed to the multiplicative processing part. Note that now the input to PreDeepRx is formed by concatenating $\mathbf{Y}$ and $\widehat{\mathbf{H}}$, instead of processing them separately.

The array $\widetilde{\mathbf{Z}}$ is then fed to the multiplicative preprocessing stage, the two key ingredients of which are 
\begin{enumerate}
\item Sparse selection of input components for multiplication;
\item Learned scaling of the imaginary part, representing a type of generalized complex conjugation.
\end{enumerate}
The former facilitates intelligent selection of inputs to multiply, while the latter allows the network to learn more easily, for example, the complex conjugation of the channel coefficients, a feature inspired by the MRC processing. The actual processing for the data $\tilde{\mathbf{z}}_{ij}$ of $ij$th RE (from $\widetilde{\mathbf{Z}}$) can then be defined with the following stages: 
\begin{itemize}
\item Expand channels with a sparse matrix $\mathbf{W}_1 \in \mathbb{C}^{M_\mathrm{ex} \times M_\mathrm{in}}$ as $\tilde{\mathbf{z}}_{ij,\mathrm{ex}}=\mathbf{W}_1 \tilde{\mathbf{z}}_{ij}$, where $M_\mathrm{ex}$ is the expanded channel count and $M_\mathrm{ex} \operatorname{mod} 3 = 0$. Note that this block essentially learns to choose which input channels to multiply.
\item Scale imaginary part of each channel by $\tilde{\mathbf{z}}_{ij,\mathrm{sc}}=\textrm{Re}\left\{ \tilde{\mathbf{z}}_{ij,\mathrm{ex}} \right\} + \mathbf{w}_2 \odot \textrm{Im} \left\{\tilde{\mathbf{z}}_{ij,\mathrm{ex}} \right\}$, where $\mathbf{w}_2 \in \mathbb{R}^{M_\mathrm{ex} \times 1 }$ and $\odot$ denotes element wise multiplication.
 \item Partition $\tilde{\mathbf{z}}_{ij,\mathrm{sc}}$ to three equal size vectors $\tilde{\mathbf{z}}_{ij,1}$, $\tilde{\mathbf{z}}_{ij,2}$, $\tilde{\mathbf{z}}_{ij,3}$.
 \item The final output of the learned preprocessing stage is then given by $\tilde{\mathbf{y}}_{ij,\mathrm{MP}}=\begin{bmatrix} \tilde{\mathbf{z}}_{ij,1} \odot \tilde{\mathbf{z}}_{ij,2} \\ \tilde{\mathbf{z}}_{ij,3} \end{bmatrix} \in\mathbb{C}^{M_\mathrm{out} \times 1}$, where $M_\mathrm{out} = \nicefrac{2}{3} M_\mathrm{ex}$.
\end{itemize}
In the above,  $\mathbf{W}_1$ and $\mathbf{w}_2$ are learned during the training procedure, and the same weights are used for all REs. Having repeated the multiplicative processing for all REs, the resulting array $\widetilde{\mathbf{Y}}_\mathrm{MP} \in \mathbb{C}^{F \times S \times M_\mathrm{out}}$ is fed to the primary DeepRx for further processing.

Finally, we note that the above procedure can also be implemented by applying complex convolutions with 1x1 filters and element wise multiplications to the whole TTI.

\subsection{Learning the Full MIMO DeepRx with Preprocessing}

\begin{table}
  \setlength{\tabcolsep}{3pt}
    \renewcommand{\arraystretch}{1.3}
    \footnotesize
    \centering
    \caption{Configurations of the proposed MIMO DeepRx architectures for the considered scenario of $N_T=4$ and $N_R = 16$. Dilations are applied only in the DeepRx blocks \cite{honkala2020}. The number of virtual layers in the MRC-based transformation is $M_T=6N_T=24$.}
    \begin{tabular}{|l|p{1.7cm}|p{1.7cm}|p{2.8cm}|}
    \hline
    \textbf{Layer} & \multicolumn{2}{l|}{\textbf{MRC-based}} & \textbf{Fully learned}\\
		\hline\hline
		Input 1  & \multicolumn{3}{c|}{RX signal: $\mathbf{Y} \in \mathbb{C}$} \\
    \hline
		Input 2  & \multicolumn{3}{c|}{Raw interpolated channel estimate: $\widehat{\mathbf{H}} \in \mathbb{C}$}  \\
		\hline
		\multirow{6}{*}{PreDeepRx}&\textbf{PreDeepRx1} & \textbf{PreDeepRx2} &\textbf{PreDeepRx}\\
		\cline{2-4}
		&  \multicolumn{3}{p{6.2cm}|}{3 Resnet blocks, 3x3 convs ($\mathbb{C}$), 64--384 channels}\\
		\cline{2-4}
		& 1x1 conv ($\mathbb{C}$), 16 channels & \multirow{3.5}{1.7cm}{Output:\newline 384 channels represented as $24\times 16$ array} &  \multirow{3.5}{2.8cm}{Output:\newline 128 channels\newline\newline} \\
		\cline{2-2}
		&  Output:\newline 16 channels &  &\\
		\hline
		\multirow{2}{*}{Transformation} & \multicolumn{2}{p{3.4cm}|}{\textbf{MRC}, output has 24 channels} & \textbf{Fully learned},\newline $M_\mathrm{in}\hspace{-1mm}=\hspace{-1mm}128$, $M_\mathrm{ex}\hspace{-1mm}=\hspace{-1mm}240$, output has 160 channels  \\
		\hline
    \multirow{4}{*}{DeepRx} & \multicolumn{3}{p{6.2cm}|}{CNN consisting of 11 ResNet blocks and depthwise-separable 2D convolutional layers ($\mathbb{R}$), following the same architecture as in \cite{honkala2020} but with quadruple channel count (although limiting the maximum number of channels to 512). }\\
		\hline
    \end{tabular}
    \label{table:architecture}
		\vspace{-5mm}
  \end{table}

Table~\ref{table:architecture} describes the overall architecture of the proposed MIMO DeepRx, including the two alternative transformations. Note that we have also experimented with shallower DeepRx MIMO architectures, which also result in reasonably high performance, although we must omit these results for brevity. The transformations and their corresponding PreDeepRx components are trained jointly with the primary DeepRx network, using the bit-level cross entropy as the loss function, where the encoded TX bit sequence represents the labels (for the expression of the loss function, please refer to \cite{honkala2020}). Note that the actual output of DeepRx represents the LLRs, which must be fed through a sigmoid-function to obtain the bit probability predictions. Moreover, the calculated cross entropy for the $q$th TTI sample, denoted by $\mathit{CE}_q \left(\bm{\theta}\right)$ where $\bm{\theta}$ is the vector of trained weights, is weighted based on the signal-to-noise ratio (SNR) of the TTI. Therefore, the final loss is given by
\begin{align}
L_q \left(\bm{\theta}\right) = \operatorname{log}_2\left(1+\mathit{snr}_q\right) \mathit{CE}_q \left(\bm{\theta}\right),
\end{align}
where $\mathit{snr}_q$ is the linear SNR of the $q$th TTI. The intuition behind using this type of weighting for the loss function is to quantify the significance of each TTI based on the achievable data rate it can support. This was observed to improve the performance of MIMO DeepRx with the higher SNRs.

In the fully learned multiplicative transformation, the sparsity requirement of the expansion matrix $\mathbf{W}_1$ should be also considered in the loss function. Therefore, $L_1$-regularization is applied to $\mathbf{W}_1$, which means that the term $\alpha\Vert\mathbf{W}_1\Vert_{L_1}$, where $\alpha$ is a regularization constant ($10^{-5}$ is used in all experiments), is added to the loss function $L_q \left(\bm{\theta}\right)$.

\section{Simulation Results}

\begin{figure*}
\vspace{-3.5mm}
	\centering
	\subfloat[MRC-based transformation]{{\includegraphics[width=0.46\linewidth]{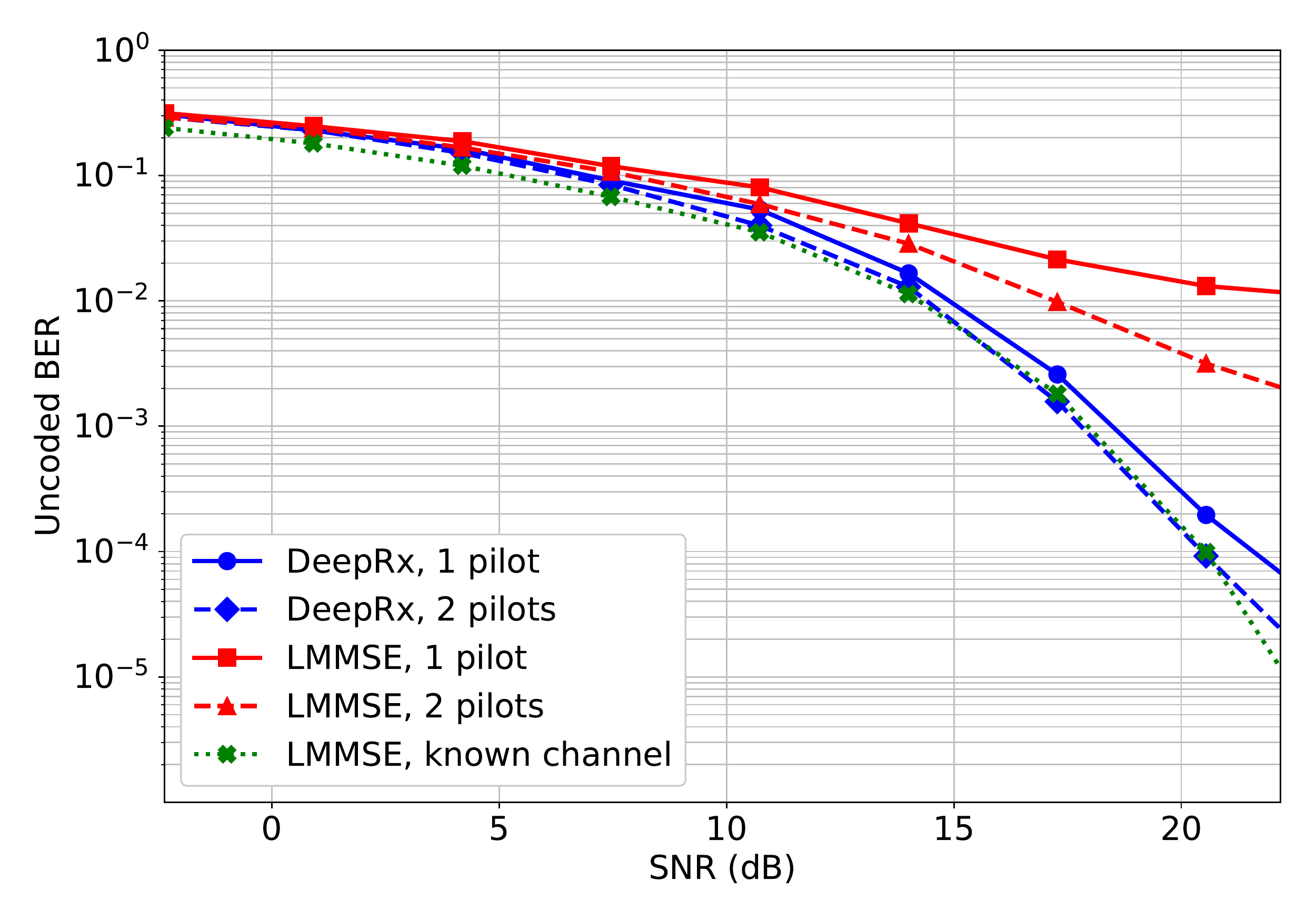}}}%
	\qquad
	\subfloat[Fully learned transformation]{{\includegraphics[width=0.46\linewidth]{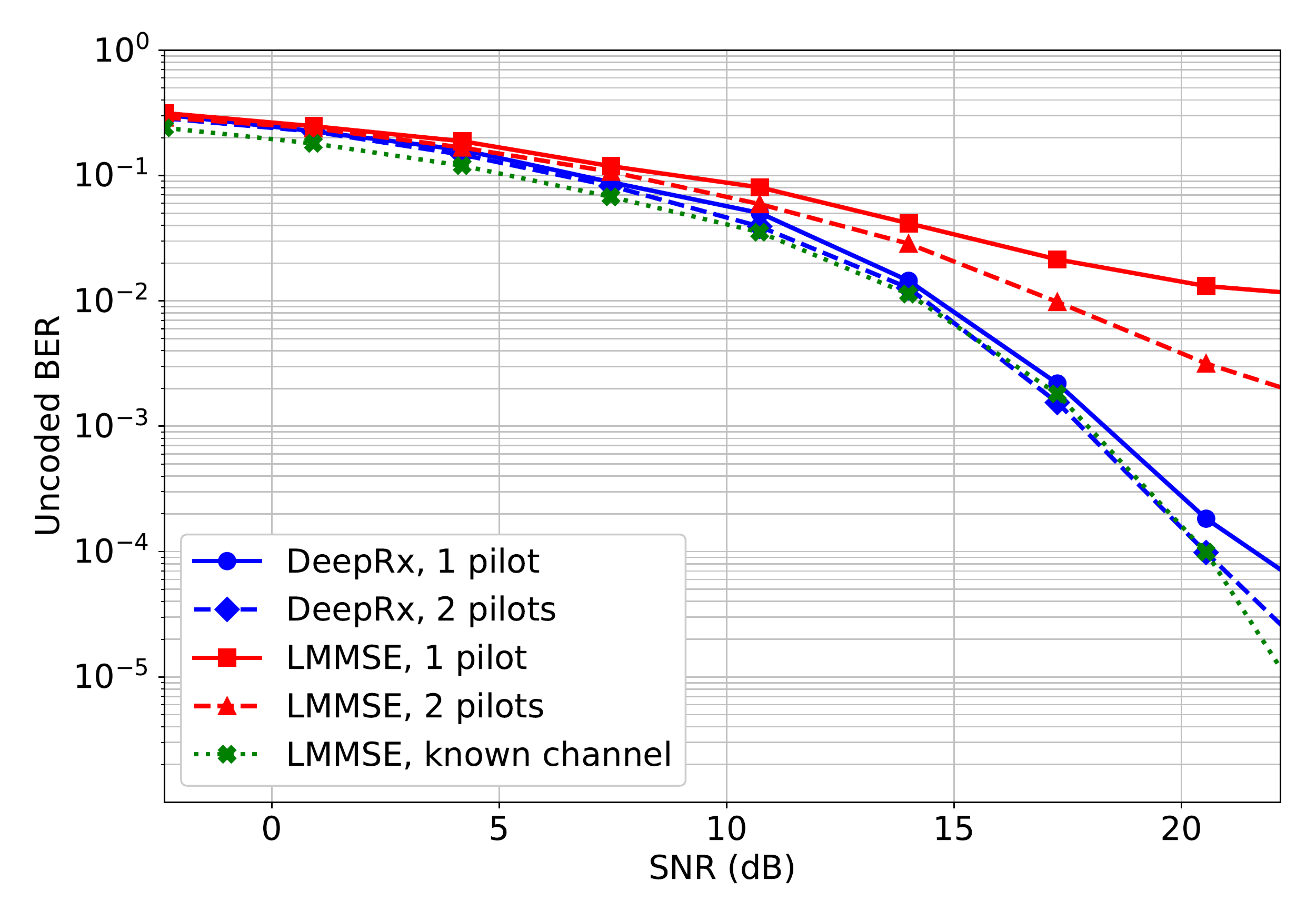}}}%
	\caption{Uncoded BER over all validation channel models for the MRC-based and fully learned transformations.}%
	\label{fig:comparison}%
	\vspace{-4mm}
\end{figure*}

Let us then evaluate the performance of the proposed MIMO DeepRx architecture using 5G simulations. The training and validation data is generated with a link-level simulator implemented with Matlab's 5G Toolbox \cite{Matlab5G}, which is modeling a 5G physical uplink shared channel (PUSCH) in a single-user-MIMO (SU-MIMO) scenario. The parameter values used in the simulations are listed in Table~\ref{table:param}. Each individual data set contains 500~000~TTIs, of which 60\% is used for training, and a subset of the remaining 40\% is used for validation. The randomization of the parameters is repeated every 10 TTIs, using the ranges and distributions indicated in Table~\ref{table:param}. Moreover, two different DMRS configurations are used: one where only the 3rd OFDM symbol of the TTI contains pilots, and another where the 3rd and 12th symbols contain pilots. In the simulations there are four transmission layers and two orthogonal pilot patterns. Hence, two layers share the same pattern via the CDM procedure. To avoid overfitting, we randomly shuffled the TX layers and corresponding ground truth bits, while keeping the RX streams untouched, each time a TTI sample was used in training.

\begin{table}
  \setlength{\tabcolsep}{2pt}
    \renewcommand{\arraystretch}{1.3}
    \footnotesize
    \centering
    \caption{Simulation parameters for training and validation.}
    \begin{tabular}{|l|p{1.8cm}|p{1.75cm}|l|}
    \hline
    \textbf{Parameter} & \textbf{Training} & \textbf{Validation} & \textbf{Randomization}\\
		\hline\hline
		Carrier frequency & \multicolumn{2}{c|}{2.6 GHz} & None\\
    \hline
		Channel model & TDL-B, TDL-C, TDL-D & TDL-A, TDL-E & Uniform\\
		\hline
		Spatial correlation & \multicolumn{2}{c|}{Low} & None\\
		\hline
		RMS delay spread & \multicolumn{2}{c|}{10 ns -- 300 ns} & Uniform\\
		\hline
		Maximum Doppler shift & \multicolumn{2}{c|}{0 Hz -- 325 Hz} & Uniform\\
		\hline
		SNR & \multicolumn{2}{c|}{$-4$ dB -- $32$ dB} & Uniform\\
		\hline
		Number of PRBs & \multicolumn{2}{c|}{26 (312 subcarriers)} & None\\
		\hline
		Subcarrier spacing & \multicolumn{2}{c|}{30 kHz} & None\\
		\hline
		OFDM symbol duration & \multicolumn{2}{c|}{35.7 $\mu$s} & None\\
		\hline
		TTI length & \multicolumn{2}{c|}{14 OFDM symbols} & None\\
		\hline
		Modulation scheme & \multicolumn{2}{c|}{16-QAM} & None\\
		\hline
		Code rate & \multicolumn{2}{c|}{$\frac{658}{1024}$} & None\\
		\hline
		Number of RX antennas & \multicolumn{2}{c|}{16} & None\\
		\hline
		Number of TX antennas & \multicolumn{2}{c|}{4} & None\\
		\hline
		Number of MIMO layers & \multicolumn{2}{c|}{4} & None\\
		\hline
		DMRS configuration & \multicolumn{2}{c|}{1 or 2 pilots with FD-CDM2} & Uniform\\
		\hline
    \end{tabular}
    \label{table:param}
		\vspace{-5mm}
  \end{table}

The proposed CNN-based MIMO DeepRx is compared to two conventional LMMSE receivers (Section~\ref{sec:lmmse}):
\begin{itemize}
\item One that performs least squares channel estimation and interpolates the channel estimate over the data symbols and subcarriers, as described in Section~\ref{sec:lmmse};
\item One that obtains the full channel information as a~priori knowledge.
\end{itemize}
The former represents a realistic benchmark and is therefore referred to as a practical LMMSE receiver, while the latter one approaches the upper bound of the achievable performance with LMMSE equalization.

The training setup, including the optimizer parameters and learning rate schedule, is the same as in \cite{honkala2020}. The only exception is that now we use 8 V100 GPUs in parallel, and train 160k iterations with a total batch size of $8 \cdot 12=96$, and a base learning rate of $3.5\cdot 10^{-3}$.

First, Fig.~\ref{fig:comparison} shows the uncoded bit error rates (BERs) before LDPC decoding over both the validation channel models (TDL-A and \mbox{TDL-E}). It can be observed that both the MRC-based and fully learned transformations allow the MIMO DeepRx to achieve essentially the same performance, clearly outperforming the LMMSE benchmark receiver. In fact, the transformation-aided MIMO DeepRx can nearly match the performance of the LMMSE receiver with perfect channel knowledge, even when the TTI contains just one pilot symbol. This indicates that the proposed MIMO DeepRx architectures are capable of remarkably accurate channel tracking and data-aided detection, similar to the SIMO DeepRx \cite{honkala2020}.

\begin{figure*}
\vspace{-3.5mm}
	\centering
	\subfloat[TDL-A: Uncoded BER]{{\includegraphics[width=0.46\linewidth]{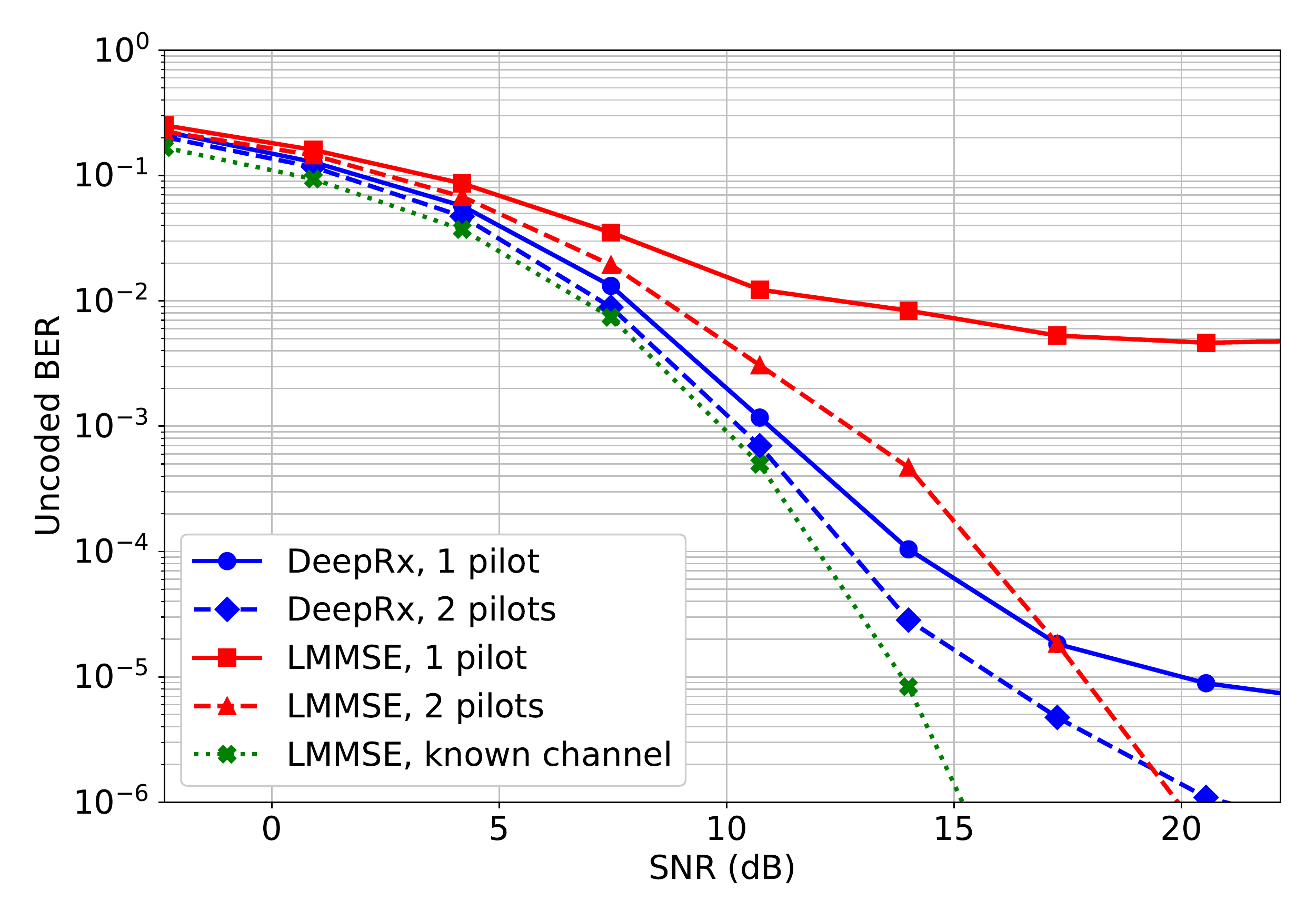}}}%
	\qquad
	\subfloat[TDL-A: Coded BER]{{\includegraphics[width=0.46\linewidth]{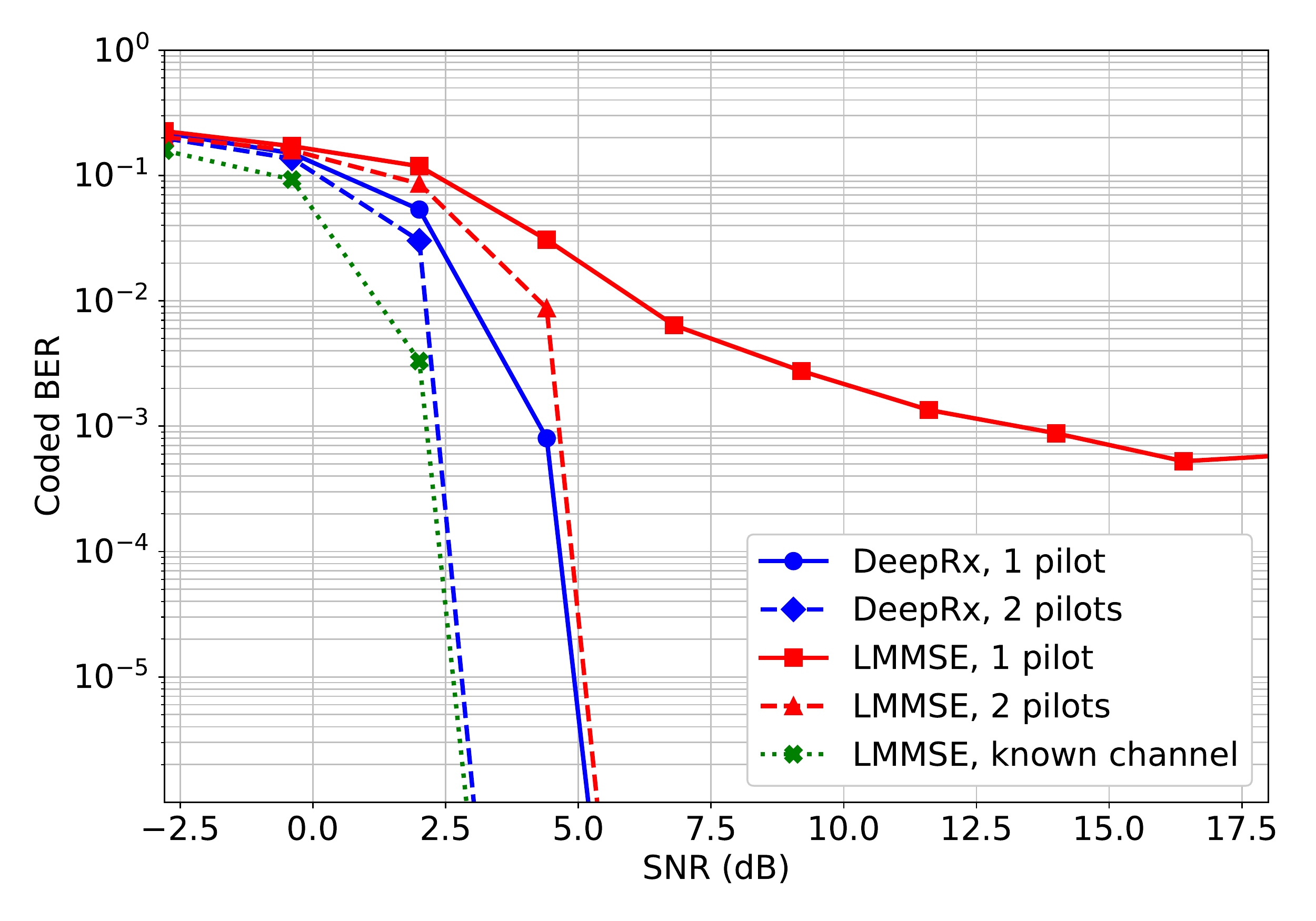}}}%
	\vspace{-3.5mm}
	\subfloat[TDL-E: Uncoded BER]{{\includegraphics[width=0.46\linewidth]{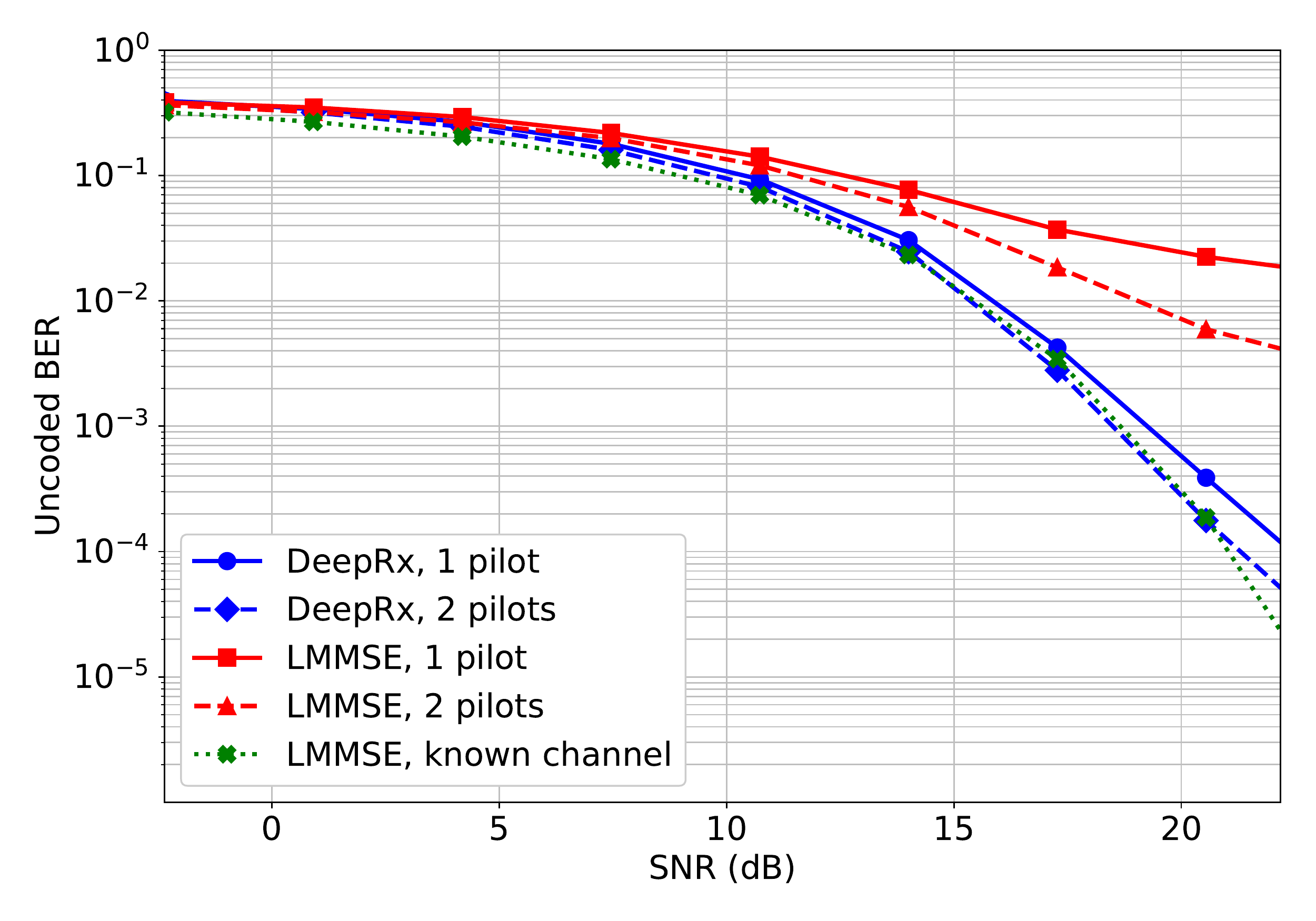}}}%
	\qquad
	\subfloat[TDL-E: Coded BER]{{\includegraphics[width=0.46\linewidth]{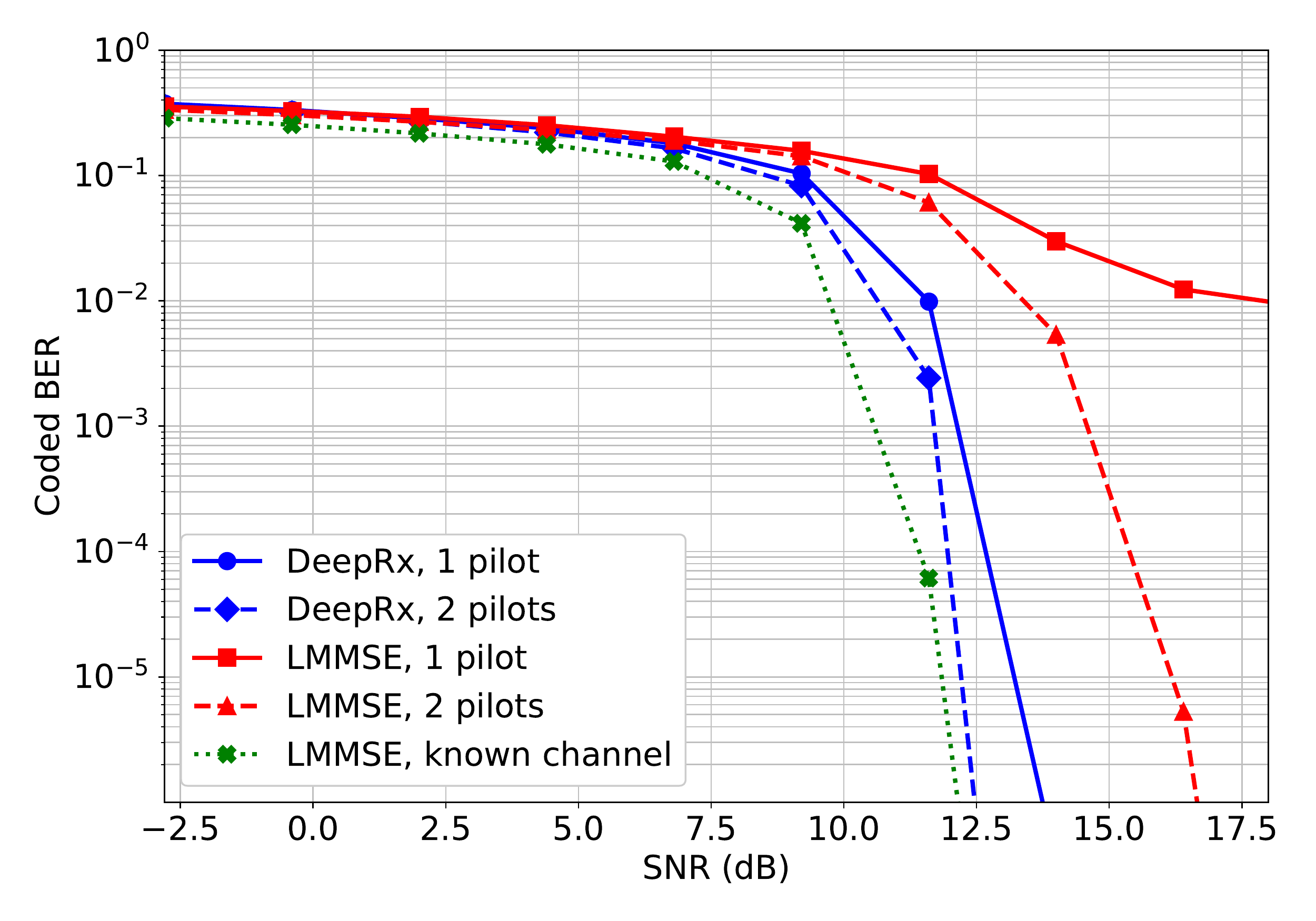}}}%
	\caption{Uncoded and coded BERs over individual channel models when utilizing the fully learned multiplicative transformation.}%
	\label{fig:fully_learned}%
	\vspace{-3.5mm}
\end{figure*}

Since both of the proposed transformation techniques were observed to provide roughly similar performance, let us next concentrate on the fully learned solution. To this end, Fig.~\ref{fig:fully_learned} shows the uncoded and coded BERs for the MIMO DeepRx utilizing the fully learned transformation. The BERs of TDL-A and TDL-E channel models are shown in separate figures in order to better understand the behavior of the MIMO DeepRx under different types of channel conditions. However, we wish to emphasize that the same model was trained to process both channels.

Investigating first Figs.~\ref{fig:fully_learned}(a)--(b) corresponding to TDL-A, it can be seen that the MIMO DeepRx with fully learned transformation outperforms the LMMSE baseline nearly over the whole SNR range. With two pilots per TTI, it can nearly match the uncoded BER of the genie-aided LMMSE up to SNRs of 14 dB, after which it seems to encounter a BER floor. However, since the waterfall region of the utilized code rate is at a relatively small SNR, the error floor of DeepRx does not impact the actual detection performance. With one pilot, the uncoded and coded BERs are somewhat higher, although the gain over the practical LMMSE with one pilot is still substantial.

Fig.~\ref{fig:fully_learned}(c)--(d) show the same results for the LOS \mbox{TDL-E} channel. In this scenario, the performance of the MIMO DeepRx with fully learned transformation is nearly on par with the LMMSE receiver having perfect channel knowledge, especially in terms of the uncoded BER. Considering the coded BER, MIMO DeepRx with just one pilot per TTI can outperform the LMMSE utilizing two pilots roughly by 2~dB. Due to the mobility, one pilot is not enough for the LMMSE receiver to even enter the waterfall region of the LDPC code. With two pilots, the MIMO DeepRx with fully learned multiplicative transformation achieves nearly the same coded BER as the genie-aided LMMSE receiver.

\section{Conclusion}

In this paper, we considered ML-based signal detection in 5G scenarios. In particular, we extended our previously presented DeepRx architecture \cite{honkala2020} to MIMO reception by proposing two alternative transformations, to be executed before the primary DeepRx part. The resulting MIMO DeepRx architectures were then trained with simulated 5G uplink data, where four layers were spatially multiplexed. The validation results showed that both of the proposed transformations allow MIMO DeepRx to achieve high performance, clearly outperforming the baseline receivers. Our future work will include further investigations into the error floor of the proposed MIMO DeepRx, which hinders its performance at very low bit error rates.

\bibliographystyle{IEEEtran}
\bibliography{IEEEabrv,references}

\begin{thebibliography}{1}
\providecommand{\url}[1]{#1}
\csname url@samestyle\endcsname
\providecommand{\newblock}{\relax}
\providecommand{\bibinfo}[2]{#2}
\providecommand{\BIBentrySTDinterwordspacing}{\spaceskip=0pt\relax}
\providecommand{\BIBentryALTinterwordstretchfactor}{4}
\providecommand{\BIBentryALTinterwordspacing}{\spaceskip=\fontdimen2\font plus
\BIBentryALTinterwordstretchfactor\fontdimen3\font minus
  \fontdimen4\font\relax}
\providecommand{\BIBforeignlanguage}[2]{{%
\expandafter\ifx\csname l@#1\endcsname\relax
\typeout{** WARNING: IEEEtran.bst: No hyphenation pattern has been}%
\typeout{** loaded for the language `#1'. Using the pattern for}%
\typeout{** the default language instead.}%
\else
\language=\csname l@#1\endcsname
\fi
#2}}
\providecommand{\BIBdecl}{\relax}
\BIBdecl

\bibitem{honkala2020}
M.~Honkala, D.~Korpi, and J.~Huttunen, ``{DeepRx}: Fully convolutional deep
  learning receiver,'' \emph{submitted to IEEE Transactions on Wireless
  Communications. arXiv preprint:1711.05101}, 2020.

\bibitem{ye18}
H.~Ye, G.~Y. Li, and B.-H. Juang, ``Power of deep learning for channel
  estimation and signal detection in {OFDM} systems,'' \emph{IEEE
  Communications Letters}, vol.~7, no.~1, pp. 114--117, 2018.

\bibitem{zhao2018}
Z.~Zhao, M.~C. Vuran, F.~Guo, and S.~Scott, ``Deep-waveform: A learned {OFDM}
  receiver based on deep complex convolutional networks,'' 2018.

\bibitem{oshea17}
T.~{O'Shea} and J.~{Hoydis}, ``An introduction to deep learning for the
  physical layer,'' \emph{IEEE Transactions on Cognitive Communications and
  Networking}, vol.~3, no.~4, pp. 563--575, Dec 2017.

\bibitem{aoudia20}
F.~A. Aoudia and J.~Hoydis, ``End-to-end learning for {OFDM}: From neural
  receivers to pilotless communication,'' \emph{arXiv preprint:2009.05261},
  2020.

\bibitem{Samuel19a}
N.~{Samuel}, T.~{Diskin}, and A.~{Wiesel}, ``Learning to detect,'' \emph{IEEE
  Transactions on Signal Processing}, vol.~67, no.~10, pp. 2554--2564, 2019.

\bibitem{pratik20}
K.~Pratik, B.~Rao, and M.~Welling, ``{RE-MIMO}: Recurrent and permutation
  equivariant neural mimo detection,'' \emph{arXiv preprint:2007.00140}, 2020.

\bibitem{shental2019}
O.~Shental and J.~Hoydis, ``Machine {LLRning}: Learning to softly demodulate,''
  in \emph{IEEE Globecom Workshops (GC Wkshps)}, Dec. 2019.

\bibitem{Matlab5G}
Mathworks, ``{Matlab 5G Toolbox},''
  \url{https://www.mathworks.com/products/5g.html}, 2020.

\end{thebibliography}

\end{document}